\documentclass{article}
\usepackage{frascatiphys}
\usepackage{graphicx}
\usepackage{amssymb}
\begin{document}
\title{ELECTROWEAK PRECISION TESTS OF THE SM}
\author{
Jens Erler        \\
{\em PRISMA$^+$ Cluster of Excellence, Johannes Gutenberg-University, 55099 Mainz, Germany}\\
on sabbatical leave from \\
{\em Departamento de F\'isica Te\'orica, IF--UNAM, 04510 CDMX, M\'exico}
}

\maketitle
\baselineskip=11.6pt
\begin{abstract}
A global survey of weak mixing angle measurements at low and high energies is presented. 
Then I will discuss theoretical uncertainties in precision observables with special emphasis on their correlations. 
The important role of vacuum polarization in global fits will also be addressed before fit results are presented.
\end{abstract}
\baselineskip=14pt

\section{Weak mixing angle, W and Higgs boson masses, and associated theory uncertainties}

I will start with a survey of measurements of the weak mixing angle, $\sin^2\theta_W$, 
as its accurate determination is becoming a global endeavor.
One can compute and measure $\sin^2\theta_W$ and relate it to the $W$ boson mass, $M_W$.
Thus, one has 3 ways of obtaining it, yielding a doubly over-constrained system at sub-per mille precision.
As this system involves relations between couplings and masses of the Standard Model (SM) particles, 
this is {\em the\/} key test of electroweak symmetry breaking.
Moreover, comparisons of measurements at different scales or between different initial or final states provide a window 
to physics beyond the SM that would remain closed with only one kind of determination, even if that would be extremely precise.

One approach to measure $\sin^2\theta_W$ is to tune to the $Z$ resonance, 
where one can measure forward-backward (FB) or left-right (LR) asymmetries
(the latter if one has at least one polarized beam) in $e^+ e^-$ annihilation around the $Z$ boson mass, $M_Z$.
Or one can reverse initial and final states and measure the FB asymmetry in $pp$ or $p\bar p$ Drell-Yan annihilation
in a larger window around $M_Z$.

A very different route is to go to lower energies, and consider purely weak processes.
Using neutrinos in the deep inelastic regime ($\nu$DIS),
where scattering occurs to first approximation off individual quarks, rates are relatively large.
Very recently the process called Coherent Elastic Neutrino Nucleus Scattering (CE$\nu$NS) as has been observed 
for the first time by the COHERENT Collaboration\cite{Akimov:2017ade} at Oak Ridge.

An alternative strategy to eliminate the electromagnetic interaction is to perform experiments in polarized and therefore
parity-violating electron scattering\cite{Erler:2014fqa} (PVES),
measuring tiny cross section asymmetries between left-handed and right-handed polarized initial states,
\begin{equation}
A_{LR} = \frac{\sigma_L - \sigma_R}{\sigma_L + \sigma_R}\ .
\end{equation}
Just as for the neutrino case, one may consider a purely leptonic process,
specifically polarized M\o ller scattering, $\vec e^- e^- \to e^- e^-$\cite{Anthony:2005pm}.
And again one can scatter deep inelastically (eDIS), but there is an important difference to $\nu$DIS.
Because of the small cross sections in $\nu$ scattering one needs large nuclei, 
which leads to complications from nuclear physics effects, while in eDIS one may use a target as small and simple as the deuteron,
as done, {\em e.g.,\/} by the PVDIS Collaboration\cite{Wang:2014bba} at JLab.
In fact, polarized eDIS was the process that established the SM\cite{Prescott:1979dh}, and 
a high-precision measurement will be possible with SoLID at the upgraded CEBAF.
The PVES analog of CE$\nu$NS on a proton target has been completed very recently by JLab's Qweak 
Collaboration\cite{Androic:2018kni} and
provided the first direct measurement of the weak charge of the proton\cite{Erler:2003yk}, $Q_W(p)$.
The future P2 experiment\cite{Becker:2018ggl} at the MESA facility at the JGU Mainz, 
will reduce the error in $Q_W(p)$ by a factor of~3, and may also run using a $^{12}$C target which is a interesting,
because it is spherical and iso-scalar and has therefore only one nuclear form factor.
Thus, $Q_W(^{12}{\rm C})$ would be easier to interpret,
especially if form factor effects can be constrained by additional run time at larger momentum transfer $Q^2$.
PVES would then be able to disentangle the weak charges of the proton and the neutron,
and consequently the effective vector couplings of the up and down quarks to the $Z$ boson.

\begin{figure}[t]
\centering
\includegraphics[width=.49\textwidth]{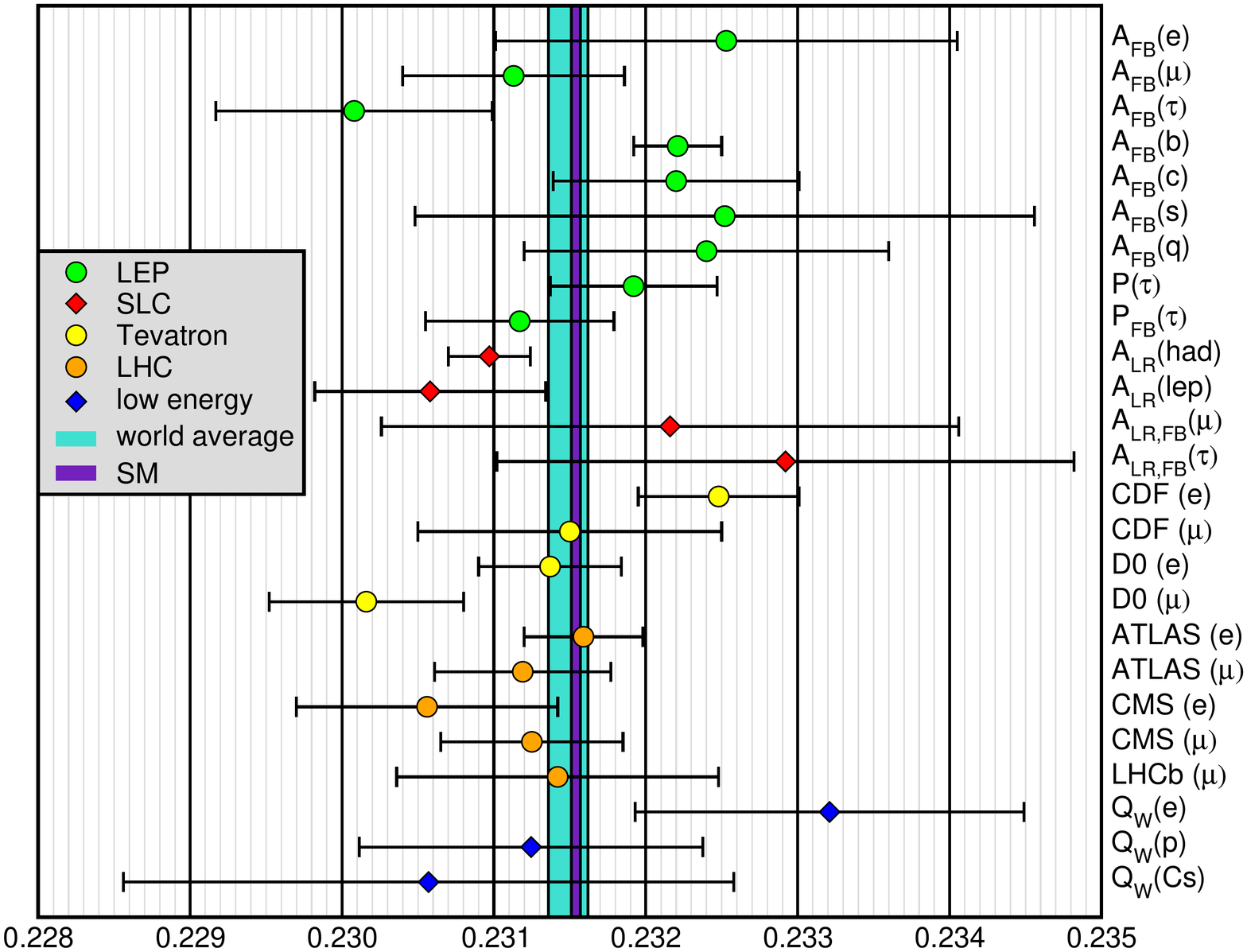}\hspace{6pt}
\includegraphics[width=.49\textwidth]{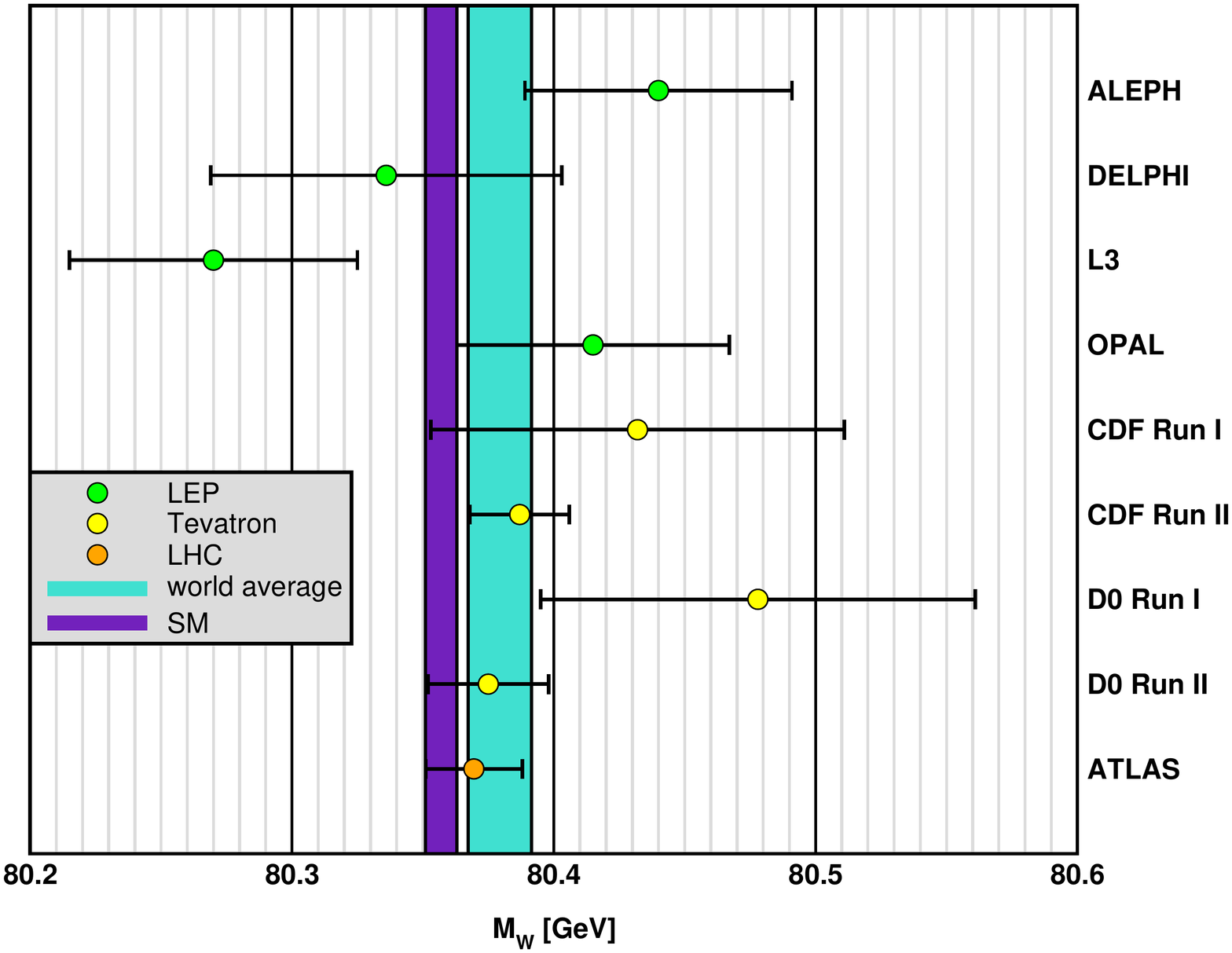}
\caption{Survey of measurements of the effective weak mixing angle (left) and the $W$ boson mass (right).}
\label{sin2thsurvey}
\end{figure}

Another newcomer are isotope ratios in atomic parity violation (APV).
Now, APV in single isotopes is a traditional way to address the weak neutral-current,
and has been studied successfully in alkali atoms\cite{Wood:1997zq}.
But one faces atomic physics complications, since one needs to understand the atomic structure
in heavy nuclei from sophisticated many-body calculations\cite{Roberts:2014bka} to a few per mille accuracy.
But most of the atomic physics effects cancel in isotope ratios.
The first such measurement has been achieved very recently at the JGU Mainz\cite{Antypas:2019raj} 
where the weak charges of Yb showed the expected isotope dependence.

Fig.~\ref{sin2thsurvey} shows the most precise determination of $\sin^2\theta_W$.
The LEP and SLC measurements in $e^+ e^-$ annihilation near the $M_Z$ pole\cite{ALEPH:2005ab}
yield the combined result, $\sin^2\theta_W = 0.23153 \pm 0.00016$.
There was a change in the extraction from the FB asymmetry for $b\bar b$ pairs at LEP, 
as the two-loop QCD correction necessary to extract the pole asymmetry 
is now known with its $b$ quark mass dependence\cite{Bernreuther:2016ccf}, 
reducing the largest LEP discrepancy with the SM by $\approx 1/4~\sigma$.
Another change affected the extraction from APV in $^{133}$Cs\cite{Wood:1997zq},
for which the Stark vector transition polarizability has been re-measured\cite{Toh:2019iro} very recently, 
shifting $|Q_W(^{133}{\rm Cs})|$ which was $1.4~\sigma$ lower than the SM value much closer to the prediction.

The leptonic FB asymmetries at the Tevatron combine to the value\cite{Aaltonen:2018dxj}
$\sin^2\theta_W = 0.23148 \pm 0.00033$.
The average\cite{Erler:2019hds} of those at the LHC, $\sin^2\theta_W = 0.23131 \pm 0.00033$, by ALTAS, CMS, and LHCb, 
assumes that the smallest theory uncertainty ($\pm 0.00025$ for ATLAS) is common to all three detectors.
Since rather different aspects of parton distribution functions are necessary for the extraction of $\sin^2\theta_W$ 
at $p\bar p$ and $pp$ colliders, the uncertainties can be assumed to be uncorrelated, 
and we find the world average, $\sin^2\theta_W = 0.23149 \pm 0.00013$,
in excellent agreement with the global fit result, $\sin^2\theta_W = 0.23153 \pm 0.00004$.

Fig.~\ref{sin2thsurvey} also shows a comparison of $M_W$ results.
In contrast to $\sin^2\theta_W$, one observes better mutual agreement among the various measurements 
at LEP\cite{Schael:2013ita}, the Tevatron\cite{Aaltonen:2013iut}, and by ATLAS\cite{Aaboud:2017svj}, but their average,
$M_W = 80.379  \pm 0.012~{\rm GeV}$, is $1.5~\sigma$ higher than the SM prediction,
$M_W = 80.361 \pm 0.005~{\rm GeV}$.

The indirect and global fit results for $M_W$ and $\sin^2\theta_W$ account not only for theory errors
but also include an implementation of theoretical correlations\cite{Erler:2019hds}.
There are various kinds of such errors entering the fits, 
where the most important ones are from unknown higher order contributions to the gauge boson self-energies.
They can be estimated by considering the expansion parameters involved, 
including various enhancement factors\cite{Erler:2019hds}.
We translate these loop factors into uncertainties in the oblique parameters\cite{Peskin:1991sw} $S = S_Z$, $T$, and 
$U = S_W - S_Z$, which have been originally introduced to parameterize potential new physics contributions to 
electroweak radiative corrections. 
Denoting these uncertainty parameters by $\Delta S_Z$, $\Delta T$ and $\Delta U$, and assuming them to be sufficiently different 
(uncorrelated) {\em induces\/} theory correlations between different observables.
We find $\Delta S_Z = \pm 0.0034$, $\Delta T = \pm 0.0073$, and $\Delta U = \pm 0.0051$.

The top quark mass determined from global fits to all data except $m_t$ from the Tevatron and LHC,
including (excluding) these uncertainties, is $m_t = 176.5 \pm 1.9~(1.8)~{\rm GeV}$.
This represents a $1.8~(1.9)~\sigma$ larger value than the direct measurement\cite{Erler:2019hds}
$m_t = 172.90 \pm 0.47~{\rm GeV}$. 
Similarly, global fits to all data except for the direct $M_H = 125.10 \pm 0.14~{\rm GeV}$ constraint\cite{Erler:2019hds} from the LHC,
give $M_H = 90^{+17}_{-15}~{\rm GeV}$ and $M_H = 91^{+18}_{-16}~{\rm GeV}$,
showing only slightly increased central value and uncertainty and reduced tension
with the directly measured value once theory uncertainties are included.

\section{Vacuum polarization in global fits}
The electromagnetic coupling at the $Z$ peak, $\alpha(M_Z)$, is needed to predict $M_W$ and $\sin^2\theta_W$.
To this end, three different groups have analyzed hadron production data in  $e^+ e^-$ annihilation, 
and in some cases $\tau$~decay spectral functions which by approximate isospin symmetry yield additional information on the former.
Or one can use perturbation theory for at least part of the calculation, and only rely on data in the hadronic region up to about
2~GeV, and then perform a renormalization group evolution\cite{Erler:2017knj} (RGE),
which depends on the strong coupling $\alpha_s$, and the charm and bottom quark $\overline{\rm MS}$ masses, 
$\hat{m}_c$ and $\hat{m}_b$.  
The results of the different approaches agree well, where for references and a discussion, I refer to Ref.\cite{Erler:2017knj}.

The data used for the hadronic part also enter other observables present in global electroweak fits,
inducing another source of uncertainty correlation.
{\em E.g.,\/} they are crucial for the SM prediction of the muon anomalous magnetic moment, $a_\mu$, 
where they enter first at two loops and generate a correlation with $\alpha(M_Z)$, 
and both are in turn anti-correlated with three-loop vacuum polarization in $a_\mu$.
Because the muon mass scale is rather low, most of the evaluation of the hadronic vacuum polarization contribution 
to $a_\mu$ is based on data.
However, there is a fraction that can be computed perturbatively. 
In particular, the heavy quark contributions are fully accessible in perturbation theory\cite{Erler:2000nx},
which for the charm contribution yields,
$a_\mu^c = (14.6 \pm 0.5_{\rm PQCD} \pm 0.2_{\hat{m}_c} \pm 0.1_{\alpha_s}) 10^{-10}$,
and where the errors are from the truncation of the perturbative series at ${\cal O}(\alpha_s^2)$, 
and the parametric errors in $\hat{m}_c(\hat{m}_c)$ and $\alpha_s$.
This in excellent agreement with the very recent lattice result in Ref.\cite{Gerardin:2019rua} and of very similar precision.
Similarly, $a_\mu^b =  0.3 \times 10^{-10}$, which has not been computed on the lattice, yet.
Note, that Ref.\cite{Gerardin:2019rua} finds a rather large total hadronic vacuum polarization contribution, so that if confirmed, there 
would cease to be a conflict between the measurement of $a_\mu$ and the SM, which currently amounts to more than $3~\sigma$.
But then there would be a new discrepancy between the dispersive and lattice gauge theory approaches to vacuum polarization.

\begin{figure}[t]
\centering
\includegraphics[width=.53\textwidth]{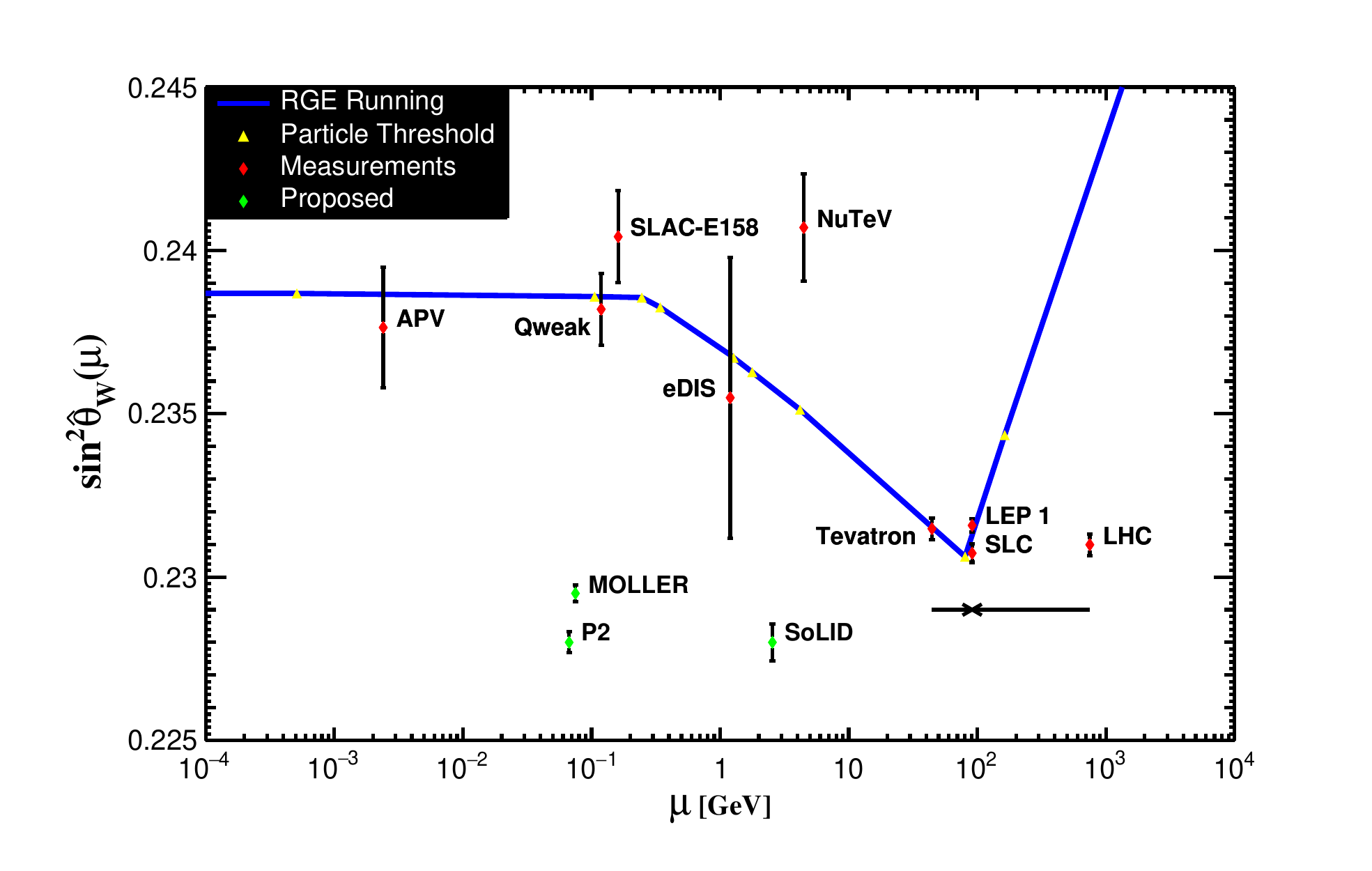}
\caption{Renormalization group evolution (running) of the weak mixing angle (updated from Ref.\cite{Erler:2017knj}).}
\label{runnings2w}
\end{figure}

$\sin^2\theta_W(0)$ enters many low-energy electroweak observables, and Fig.~\ref{runnings2w} shows
that future low-energy PVES experiments will be at the precision level of the LEP and SLC measurements.
To compute the RGE in the non-perturbative region, one needs the same kind of data that enters the calculation of $\alpha(M_Z)$. 
This part needs to be subdivided into two pieces because the vector couplings of the $Z$ boson differ from the electric charges,
implying that there is a piece that is not directly related to $\alpha(M_Z)$ and necessitating a study of the effect and
uncertainty associated with the corresponding flavor separation.
Estimates of the singlet piece and isospin breaking effects are also required.
The overall uncertainty is negligible compared to any upcoming low-energy determination of $\sin^2\theta_W$ 
in the foreseeable future\cite{Erler:2017knj}.

The final application of vacuum polarization are heavy quark mass determinations.
If one employs as input quantities only the electronic decay widths of the narrow resonances, and compares two different moments 
of the relevant vacuum polarization function, one obtains simultaneous information on the quark mass and the continuum contribution.
The constraint on the latter can then be compared with the experimental determination of electro-production of the open heavy quark.
This results in an over-constrained system, where any residual difference can be taken as an error
estimate\cite{Erler:2016atg} of non-perturbative effects which are supposedly small but possibly not entirely negligible. 
This strategy has been applied to $\hat{m}_c$ resulting in the precision determination\cite{Erler:2016atg},
$\hat{m}_c(\hat{m}_c) = 1272 \pm 8 + 2616 [\alpha_s(M_Z) - 0.1182]~{\rm MeV}$,
where the central value is in very good agreement with recent lattice results\cite{Aoki:2019cca} and of comparable precision.

\section{Results and conclusions}
A simple example to illustrate how global fits constrain physics beyond the SM
is the $\rho_0$ fit, where one assumes that the new physics is mainly affecting the $\rho$~parameter,
quantifying the neutral-to-charged current interaction strengths.
{\em E.g.,\/} any electroweak doublet with a mass splitting, $\Delta m^2 \geq (m_1 - m_2)^2$, 
contributes to $\rho_0$ positive definitely.
It might appear that there is no decoupling, so that even a doublet with Planck scale masses 
but electroweak size splitting may give observable effects in experiments at much lower energies,
but this is not the case, as there is a see-saw type suppression of $\Delta m^2$ in any given model.
Indeed, the leading contributors to $\rho_0$ in the SM effective field theory are dimension 6 operators, 
so that these effects are suppressed by at least two powers of the scale of new physics.
The global fit yields\cite{Erler:2018pdg} $\rho_0 =  1.00039 \pm 0.00019$,
which is 2~$\sigma$ higher than the SM value, $\rho_0 \equiv 1$, and a manifestation of the tension in $M_W$ discussed earlier.   
It is amusing to point out that at face value, one even finds a non-trivial 95\%~CL {\em lower\/} bound 
on the sum of all such mass splittings.
This strongly disfavors, {\em e.g.,\/}  zero hypercharge, $Y = 0$, Higgs triplets for which $\rho_0 < 1$.
On the other hand, a Higgs triplet with $|Y| = 1$ is consistent with the data provided its vacuum expectation value
is around 1\% of that of the SM doublet.

\begin{figure}[t]
\centering
\includegraphics[width=.53\textwidth]{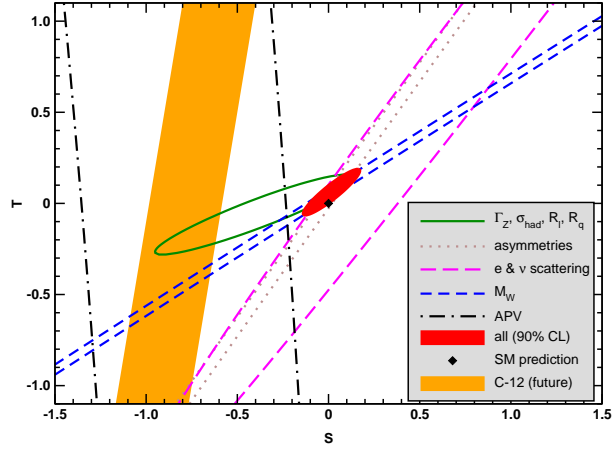}
\caption{$T$ {\em vs.\/} $S$ for various data sets.  
Also shown is the impact that the $^{12}$C PVES measurement would have if it could be performed with a relative error of $0.3\%$. 
This yields a different slope in the $ST$-plane.}
\label{STplot}
\end{figure}

Another example is a fit\cite{Erler:2018pdg} to the $S$ and $T$ parameters\cite{Peskin:1991sw}, 
$S = 0.02 \pm 0.07$ and $T = 0.06 \pm 0.06$ with a correlation of 81\%,.
It is illustrated in Fig.~\ref{STplot}. 
$U = 0$ is fixed, as it is generally suppressed by 2~extra factors
of the new physics scale\cite{Grinstein:1991cd} compared to $S$ and $T$.
Remarkably, with these 2~extra degrees of freedom, the minimum $\chi^2$ drops by 4.2~units.
One can interpret the $S$ and $T$ parameters in a variety of new physics models,
if one assumes that non-oblique effects are absent or small.
{\em E.g.,\/} the mass of the lightest Kaluza-Klein state\cite{Carena:2006bn} in warped extra dimensions\cite{Randall:1999ee}
should satisfy the bound $M_{KK} \gtrsim 3.2$~TeV, while the lightest vector state 
in minimal composite Higgs models\cite{Pich:2013fea} is bound by $M_V \gtrsim 4$~TeV\cite{Erler:2018pdg}. 

To conclude, both, the LHC and low-energy measurements are approaching LEP and SLC precision in $\sin^2\theta_W$.
There are new players represented by COHERENT\cite{Akimov:2017ade}, Qweak\cite{Androic:2018kni}, 
and APV isotope ratios\cite{Antypas:2019raj}, where with the lower precision of these first measurements,
it is currently more interesting to assume the validity of the SM, and to use them to constrain neutron skins
(the difference of the neutron and proton radii in nuclei), or more generally form factor effects.

\section*{Acknowledgements}
\noindent
This work is supported by CONACyT (M\'exico) project 252167--F,
the German--Mexican research grant SP 778/4--1 (DFG) and 278017 (CONACyT), and by PASPA (DGAPA--UNAM).
I gratefully acknowledge the hospitality and support by 
the Helmholtz-Institute and the THEP group at the Institute of Physics at the JGU Mainz.
Finally, I thank Rodolfo Ferro-Hern\'andez for updating Fig.~\ref{runnings2w}, and
I am indebted to Werner Bernreuther and Long Chen for a dedicated update and a private communication 
w.r.t.\ Ref.\cite{Bernreuther:2016ccf}.

\end{document}